\documentclass[aps,prl,twocolumn,groupedaddress]{revtex4}

\usepackage{subfigure}
\usepackage {graphicx}

\newcommand {\Mg} {Mg$^{\mathrm {2+}}$}
\newcommand {\Na} {Na$^{\mathrm {+}}$}

\begin{document}

%\preprint{}

\title{Persistence Length Changes Dramatically as RNA Folds}


% \affiliation can be followed by \email, \  homepage, \thanks as well.
%\author{Gokhan Caliskan$^{1,2}$, Changbong Hyeon$^3$, Ursula Perez-Salas$^{4,5}$, Robert M. Briber$^6$, Sarah A. Woodson$^7$, and D. Thirumalai$^3$\footnote[1]{To whom correspondence should be addressed. Phone: 301-405-4803; Fax: 301-314-9404; E-mail: thirum@glue.umd.edu}}
\author{G. Caliskan$^{1,2}$, C. Hyeon$^3$, U. Perez-Salas$^{4,5}$, R. M. Briber$^6$, S. A. Woodson$^7$, and D. Thirumalai$^3$}
\affiliation{$^1$T. C. Jenkins Department of Biophysics, Johns Hopkins University, 3400 N. Charles St., Baltimore, MD  21218\\
$^2$Center for Neutron Research, National Institute of Standards and Technology, Gaithersburg, MD 20899\\
$^3$Biophysics Program, Institute For Physical Science and Technology, University of Maryland, College Park, MD 20742\\ 
$^4$Center for Neutron Research, National Institute of Standards and Technology, Gaithersburg, MD 20899 \\
$^5$Department of Physiology and Biophysics, University of California at Irvine, Irvine, CA 92697 \\
$^6$Department of Materials Science and Engineering, University of Maryland, College Park, MD 20742 \\
$^7$T. C. Jenkins Department of Biophysics, Johns Hopkins University, 3400 N. Charles St., Baltimore, MD  21218}
\date{\today}

\begin{abstract}
We determine the persistence length, $l_p$, for a bacterial group I ribozyme as a function of concentration of monovalent and divalent cations 
by fitting the distance distribution functions $P(r)$ obtained from small angle X-ray scattering  intensity data to the 
asymptotic form of the calculated $P_{WLC}(r)$ for a worm-like chain (WLC).  The $l_p$ values change 
dramatically over a narrow range of \Mg~ concentration from $\sim$21 \AA~ in the unfolded state (\textbf{U}) to $\sim$10 \AA~ in the compact ($\mathrm{I_C}$)
and native states.  
Variations in $l_p$ with increasing \Na~ concentration are more gradual.
In accord with the predictions of polyelectrolyte theory we find $l_p \propto 1/ \kappa^2$ where $\kappa$ is 
the inverse Debye-screening length.  
\end{abstract}

%\keywords{}

\maketitle


Elucidating the mechanisms by which RNA molecules self-assemble to form three dimensional structures is a challenging problem \cite{Treiber99COSB, Sosnick03COSB, Thirumalai05BioChem, Brion97AnnRevPhysChem}.  
Because the native state (\textbf{N}) cannot form without significantly neutralizing the 
negative charge on \cite{Misra98Biopolymers, Thirumalai01AnnRevPhysChem} the  phosphate group, RNA 
folding is sensitive to the valence, size, and shape of the counterions.  At low counterion 
concentrations ($C$) RNA is unfolded (\textbf{U}) in the sense that it contains isolated stretches of base-paired \textit{stem loops} that have
large dynamical fluctuations.  
When $C > C_m$, the midpoint of the transition from \textbf{U} to the \textbf{N}, RNA becomes compact as a result of formation 
of tertiary contacts.  For many RNA molecules, such as the \textit{Tetrahymena ribozyme} and RNase P, 
folding to the native state
is preceded by the formation of multiple metastable kinetic intermediates (\textbf{I}) \cite{Treiber99COSB,Sosnick03COSB,Pan97JMB}.

The large  dynamic conformational
fluctuations in the \textbf{U} and \textbf{I} states make it difficult to characterize their structures. However,
small angle scattering experiments can be used to determine the shape of RNA as it folds. 
The conformation of RNA in the \textbf{U}, \textbf{N}, and the \textbf{I} states is characterized by $R_g$, the radius of gyration, and $l_p$, 
the persistence length.  
Small Angle X-ray Scattering (SAXS)
\cite{Takamoto04JMB, Sosnick03COSB,Fang00Biochem,Russell00NSB,Russell02PNAS} and 
Small Angle Neutron Scattering (SANS) \cite{Perez-Salas04Biochem}
experiments have been used to obtain $R_g$ as a function of counterions for a number of RNA molecules. 
In contrast, $l_p$, which is a function of $C$ and valence and shape of counterions, is
more difficult to obtain.  

In this letter, we use SAXS data and theoretical results for the worm-like chain (WLC) to
obtain  $l_p$ for a 195 nucleotide group I ribozyme from pre-tRNA(Ile) of the \emph{Azoarcus} bacterium as a function of $C$ for 
monovalent and divalent counterions.  
The major conclusions of the present study are:
(i) The experimentally determined distance distribution functions $P(r)$ can be accurately fit 
using the theoretical results for worm-like chains for $r/R_g > 1$ where $R_g$ is the radius of gyration of RNA.   The $l_p$ values, 
which were calculated by fitting  $P(r)$ to $P_{WLC}(r)$ for $r > R_g$, change dramatically from $l_p \simeq 21$ \AA~ in the \textbf{U} state to $l_p \simeq 10$ \AA~ 
in the compact conformation. 
(ii) The large reduction in $l_p$ occurs abruptly over a narrow concentration range in \Mg~ whereas the decrease of $l_p$ in \Na~ is gradual. This result suggests
that the compaction of RNA resembles a first order transition in the presence of multivalent counterions.  
(iii) For both \Na~ and \Mg , the persistence length scales as $l_p \simeq l_D^2$ where $l_D$ is the Debye-screening length. 
From this finding, which is in accord with the predictions of polyelectrolyte theory, we find that the intrinsic persistence length of RNA is $l_p^o \simeq 10$ \AA~.  

The \emph{Azoarcus} ribozyme was transcribed \textit{in vitro} as described previously \cite {Rangan03PNAS}.  We carried out SAXS measurements at Argonne National Lab 
Advanced Photon Source (BIOCAT) beamline using 1.05 \AA~ X-rays that corresponds to 11.8keV in energy.  A sample to detector distance of 1.89m allowed us to probe 
 momentum transfer ($Q$) in the range from ($\sim$0.007 to $\sim$0.266)\AA $\mathrm{^{-1}}$.  A quartz capillary flow cell was used to 
minimize the radiation damage due to X-ray exposure of a given RNA chain \cite{Fang00Biochem}.  The measurements at various 
flow rates showed that X-ray radiation damage is negligible.  
Each measurement was  averaged from four separate exposures of two seconds each.  The SAXS profiles were corrected for the 
background signal which was measured at same buffer concentrations in the absence of RNA.  Using the background corrected SAXS intensity as a function of $Q$, ($I(Q)$) 
(Fig. (1A)) the distance distribution function, $P(r)$,
was calculated by an indirect inversion algorithm \cite{Semenyuk05web}.  The square of the radius of 
gyration is given by $R_g^2 = \int r^2P(r) \mathrm {d}r/\int 2P(r)\mathrm{d}r$.
\begin{figure}
  \includegraphics[width=2.5in]{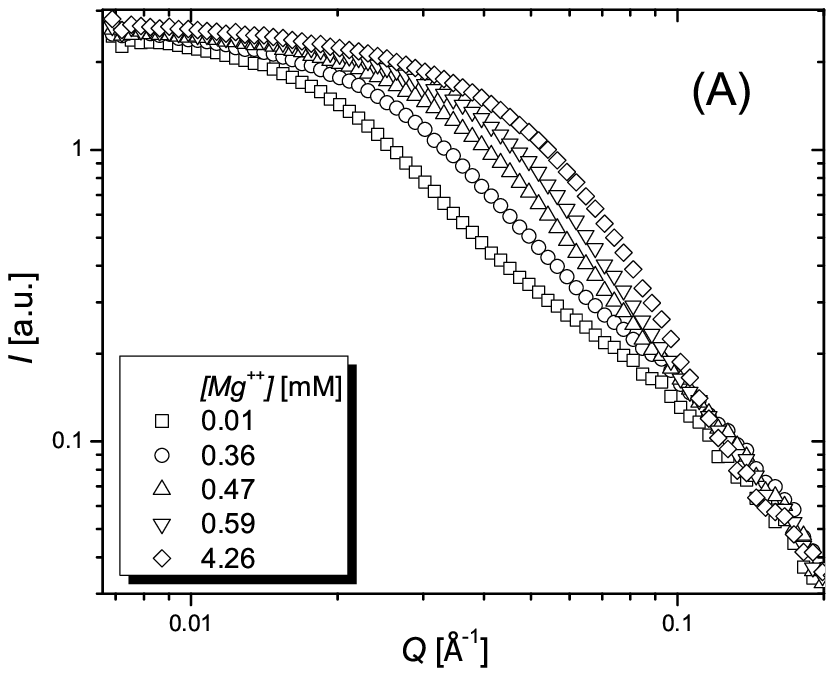}
  \includegraphics[width=2.5in]{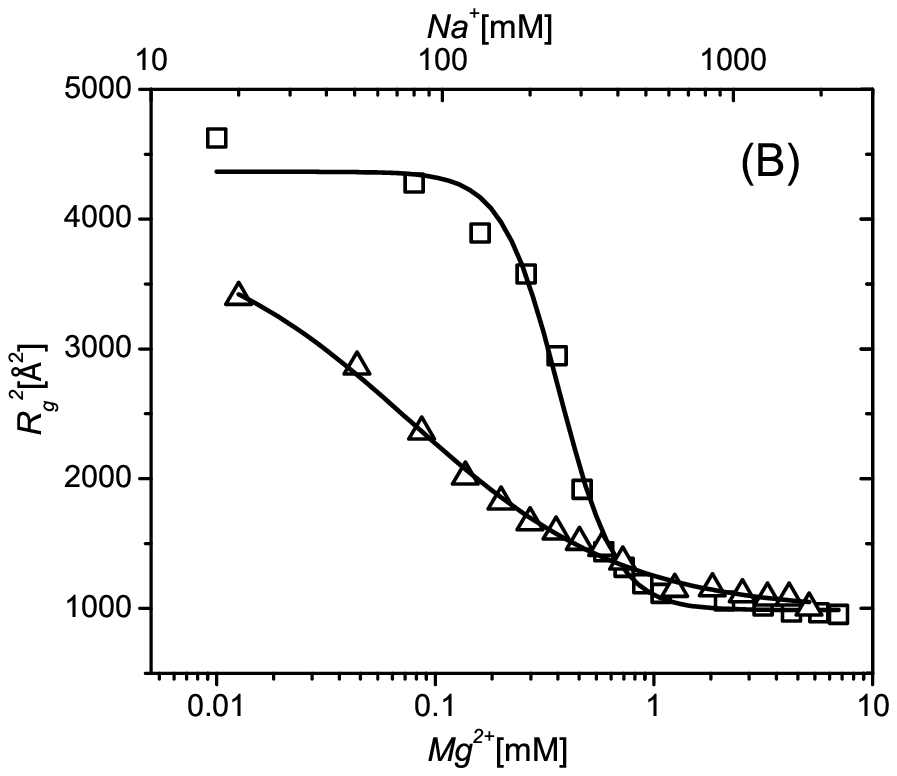}
 \caption {(A) Scattering intensity  $I(Q)$ as a function of $Q$ for 195 nucleotide \emph{Azoarcus} ribozyme at different as a function
of \Mg~ concentration in 20mM Tris buffer,  pH = 7.5 at $32^oC$. The values of \Mg~ concentrations are given in the inset. 
(B) The dependance of $R_g^2$ on \Mg~ (squares) and $Na^+$ (triangles) concentration. Solid line is the fit using Hill equation in  the form 
$A(1- [Mg^ {2+} ]^n/(C_m^n + [ Mg^{2+}]^n)) + y_o$, where $A, n$ and $y_o$ are adjustable parameters. 
We find $n$ is 3.33 and 1.20 for \Mg~ and \Na, respectively.}
\end{figure}

Single molecule measurements of RNA subject to tension and our analysis of PDB structure have shown that the force-extension curves can be fit 
using a worm-like chain (WLC) model \cite{Liphardt01Science}. Based  on these studies we assume that RNA is a WLC for 
which $P(r)$ cannot be calculated analytically.   However, a simple theoretical expression has been derived \cite{Thirumalai98book} for 
the end-to-end $R_E$ distribution using a mean-field model of WLC.  
We expect that asymptotically ($r >> R_g$) the behavior of $P(r)$ and the distribution of $R_g$ or $R_E$ should have the
same functional form.  
Thus, for large $r$ we predict $P(r)$ should decay as \cite{Thirumalai98book}
\begin {equation}
P_{WLC}(r) \sim \exp \left( - \frac {1}{1-x^2} \right)
\end {equation}

\noindent where $x = l_p r/R_g^2$.
  
At all concentrations of \Na~ and \Mg, Eq. (3) fits 
the data extremely accurately as long as $r/R_g > 1$  (Fig. (2)).  The excellent fits in Fig. (2) allows us to determine $l_p$ as a function of the counterion 
concentration.  When RNA is unfolded at low \Na~ or \Mg~ concentration, $l_p \simeq$ 21 \AA~ with $R_g \simeq$ 65 \AA~.  As the concentration of \Na~ increases from about 
(20 - 200) mM, $l_p$ gradually decreases.  There is a sharp decrease in $R_g$ when [\Na ] $ \simeq$ 250 mM that is accompanied by a large reduction in the persistence length 
to $l_p \simeq$ 10 \AA.  The changes in $l_p$ are even more dramatic in \Mg~ (Fig. (2B)).  As \Mg~ increases from 0.01 mM to 0.26 mM the persistence length changes 
only by about 3 \AA~ from $l_p \simeq$21 \AA~ (0.01 mM) to $l_p \simeq $ 18.3 \AA~ (0.26 mM).  In this concentration range $R_g$ decreases from 65 \AA~ to 60 \AA.  A further 
increase in \Mg~ to 4.26 mM leads to a reduction in $R_g$ to about 31 \AA~ with a dramatic decrease in  $l_p$ to about 10 \AA.  
The \emph{near discontinuous} change in $R_g$ in \Mg~ (Fig. (1B)) suggests  a first-order coil-globule transition in \Mg. While less common in
neutral homopolymers, a  discontinuous coil-globule transition has been predicted to occur in strongly charged polyelectrolytes \cite{Ha92PhysRevA}.
\begin {figure}
 \includegraphics[width=2.5in]{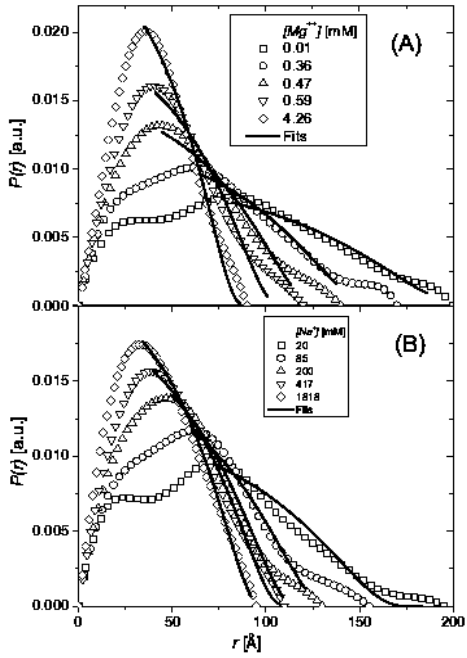}
 \caption{Distance distribution functions. (A) $P(r)$ functions at various \Mg~ concentrations at 32$^o$C are obtained using  
Eq. (1) with $Q_{min} = 0.07$ \AA~ and $Q_{max} = 0.1$ \AA. The solid lines are fits of $P(r)$ to Eq.(3) (B) Same as (A) 
except the counterion is \Na.  The concentrations of counterions are given in the insets.}
\end{figure}

To complement the experimental studies we calculated the $P(r)$ functions for the native three dimensional structure  of \textit{Azoarcus} ribozyme 
using the coordinates from X-ray crystallography crystal structure \cite{Adams04Nature} (PDB id: 1U6B) and a model based on sequence comparison \cite{Rangan03PNAS}.  
The computations  were done using the coordinates of only the heavy atoms (C, O, P, and N). 
To compare the results obtained from crystal structures  and SAXS data,  
we used only the heavy atom coordinates for chain B (excluding nucleotides 1 and 197) from 1U6B structure to compute $P(r)$. Similarly, the exon fragments were excluded from the Westhof model. 

The $P(r)$ function from the SAXS data for the  \textbf{N} state and those obtained using the X-ray structure and 
the Westhof model are in good agreement with each other and the SAXS data(Fig. 3A). The radii of gyration for the native state calculated using
$(R_g^N)^2 = \frac {1}{2N^2} \sum_i {\sum_j {(r_i -r_j)^2}}$
for the X-ray structure and the Westhof model are 31.1 \AA~ and 30.7 \AA~ respectively. 
These values agree well with the results from the SAXS data ($R_g^N$ = 30.9 \AA ). 
The $l_p$ for the native state obtained by fitting  the crystal structure $P(r)$ to Eq. (3) is 11 \AA, while 
for the Westhof model we obtain $l_p \approx$ 10.8 \AA. 
The good agreement between the crystal structure (or the Westhof model)
and the SAXS measurement for $P(r)$ and $l_p$ in the $\mathrm{I_C}$ state suggests that the effects of complexation and interparticle interactions
in the SAXS experiments are negligible.  
\begin {figure}
 \includegraphics[width=2.3in]{SAXS_PDB_comp_bw}
 \includegraphics[width=2.5in]{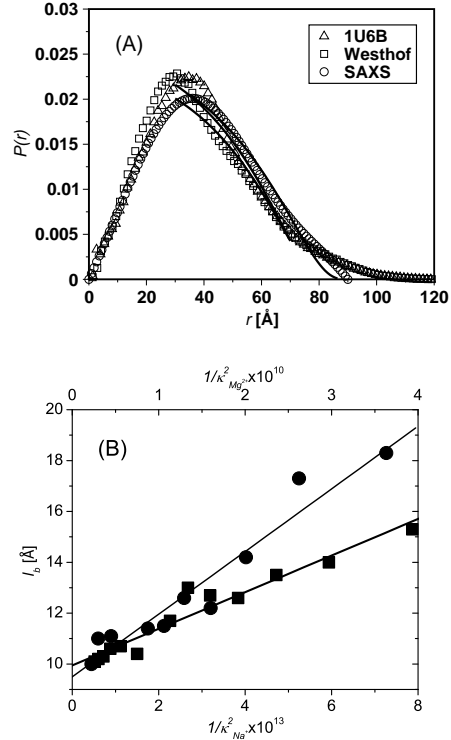}
 \caption{(A) Calculation of $P(r)$ using the coordinates of the heavy atoms from the
chain B of the crystal structure (1U6B), the Westhof model \cite{Rangan03PNAS}, and SAXS data.  The symbols are given in the inset. The solid  lines show the
fits using $P_{WLC}(r)$ in the range 30 \AA~ to 70 \AA~. In this range the root mean square
deviation of the $P(r)$ for the 1U6B from $P_{WLC}(r)$ is 0.13 $\mathrm{\AA^{-1}}$ and for the Westhof model \cite{Rangan03PNAS}
it is 0.10  $\mathrm{\AA^{-1}}$. The correlation coefficient of the fits of $P(r)$ to $P_{WLC}(r)$ is 0.98.    (B) Dependence of $l_p$ on 1/ $\kappa^2$ in \Mg~ 
(solid circles) and in \Na~ (solid squares). Lines represent fits 
to the data. Note that the $1/\kappa^2$ scale for \Mg~ is given on top.}
 \end{figure}

Since RNA appears to be a charged worm-like polyelectrolyte, it is of interest to 
ascertain if the dependance  of $l_p$  on the Debye-screening length conforms to the theoretical predictions
\cite{Odijk77JPolySciPartB,Skolnick77Macromolecules}. The dependence of $l_p$ on the 
square of the Debye-screening length ($l_D^{-2} = \kappa^2 = 8 \pi l_B I$ where $l_B =  {e^2} / {4 \pi \epsilon k_B T}$ is 
the Bjerrum length and $I$ is the ionic strength) is linear for both \Na~ and \Mg~ (Fig. (3B)). 
For both flexible \cite{Ha99JCP, Netz99EurPhysJ} and stiff polyelectrolytes \cite{Odijk77JPolySciPartB, Skolnick77Macromolecules} 
it has been shown that $l_p = l_p^o + l_p^{el}$ where $l_p^o$ is the intrinsic 
persistence length and the electrostatic contribution  is $l_p^{el} \propto 1/ \kappa^2$. 
Deviation from the OSF predictions can occur for finite-sized flexible polyelectrolytes. However, we do not expect such deviations because RNA is intrinsically stiff.
Surprisingly, over the range of \Na~ and \Mg~ concentrations in which the   \emph{Azoarcus} ribozyme undergoes the  \textbf{U} $\rightarrow$ 
$\mathrm{I_C}$ transition,  the experimental data confirm the predictions of polyelectrolyte theory.  From the linear fits of $l_p$ to
$\kappa^{-2}$ (Fig. (3B)) we 
obtain $l_p^o \simeq $ 10 \AA~ which is similar to those found for single stranded  DNA \cite{Rivetti98JMB,Tinland97Macromolecules}.

To assess if $P(r)$ for WLC can be used to fit scattering measurements on other RNA molecules we used Eq. (3) and SAXS data for RNase P \cite {Fang00Biochem} 
as a function of \Mg~ concentration.  Unlike the \emph {Azoarcus} ribozyme,  folding of RNase P is best described using three states, namely, \textbf{U}, 
an intermediate \textbf{I}, and the native state, \textbf{N} \cite{Fang00Biochem}.  The \textbf{I} state is populated in the \Mg~ range $0.02 <$ \Mg $< 0.2$ . 
From the accurate fit of the SAXS data using Eq. (3) for $r/R_g > 1$, 
the $l_p$ values are found to be 24.5\AA, 14.1\AA, and 11.6\AA~ in the \textbf{U}, \textbf{I}, and \textbf{N} states respectively.  
The largest decrease in $l_p$ and the associated $R_g$ occurs in the \textbf{U} $\rightleftharpoons $ \textbf{I} transition, 
which is consistent with the notion that the early event in RNA collapse is initiated by counterion condensation \cite{Thirumalai05BioChem}. 
 \begin {figure}
 \includegraphics[width=2.5in]{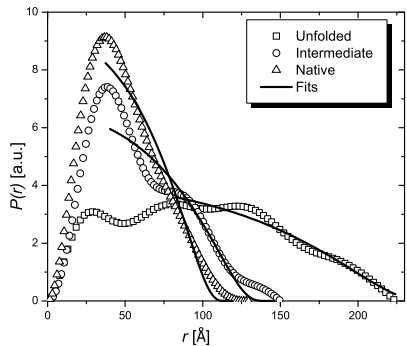}
 \caption{$P(r)$ for RNase P  from reference \cite{Fang00Biochem}.  The distance distribution function was calculated from $I(q)$ 
using Eq. (1) with a different numerical Fourier Transform method\cite{Svergun00web}. 
The \textbf{U}, \textbf{I} and \textbf{N} states are for \Mg~ concentrations 0, 0.1 , and 10 mM respectively. 
The lines are fits using Eq. (3).}
 \end{figure}
 
The present work shows that the size and flexibility of RNA molecules as a function of counterion concentration 
can be obtained using scattering experiments and the WLC model. Given that RNA is a highly branched and charged polymer, it is surprising that the 
distance distribution  functions can be described using elasticity-based polymer models for $r/R_g > 1$. Although the structural basis for 
such a behavior is not obvious, the demonstration that single stranded DNA \cite{Tinland97Macromolecules}, double stranded DNA \cite{Smith92Science},
and polypeptide chains \cite{Schuler05PNAS} also behave like WLC suggests that for compatible interactions between biomolecules the local flexibility
should be similar.

We are grateful to T. R. Sosnick for providing the $P(r)$ data for RNase P in tabular form. This work was supported 
in part by a grant from the National Science Foundation to DT (grant number 05-14056) and the National Institutes of Health to SAW. 
\bibliography{bibliography}
\end{document}